\documentclass[12pt,preprint]{aastex}
\usepackage{pslatex}
\slugcomment{Accepted for publication in the Astronomical Journal}

\begin{document}
\title{{\it HST/ACS} Coronagraphic Imaging of the AU Microscopii Debris Disk}
\author{John E. Krist\altaffilmark{1},
D.R. Ardila\altaffilmark{2}
D.A. Golimowski\altaffilmark{2},
M. Clampin\altaffilmark{7},
H.C. Ford\altaffilmark{2},
G.D. Illingworth\altaffilmark{3},
G.F. Hartig\altaffilmark{1},
F. Bartko\altaffilmark{4}, 
N. Ben\'{\i}tez\altaffilmark{2},
J.P. Blakeslee\altaffilmark{2},
R.J. Bouwens\altaffilmark{3},
L.D. Bradley\altaffilmark{2},
T.J. Broadhurst\altaffilmark{5},
R.A. Brown\altaffilmark{1},
C.J. Burrows\altaffilmark{},
E.S. Cheng\altaffilmark{6},
N.J.G. Cross\altaffilmark{2},
R. Demarco\altaffilmark{2},
P.D. Feldman\altaffilmark{2},
M. Franx\altaffilmark{8},
T. Goto\altaffilmark{2},
C. Gronwall\altaffilmark{9},
B. Holden\altaffilmark{3},
N. Homeier\altaffilmark{2},
L. Infante\altaffilmark{10}
R.A. Kimble\altaffilmark{7},
M.P. Lesser\altaffilmark{11},
A.R. Martel\altaffilmark{2},
S. Mei\altaffilmark{2},
F. Menanteau\altaffilmark{2},
G.R. Meurer\altaffilmark{2},
G.K. Miley\altaffilmark{8},
V. Motta\altaffilmark{10},
M. Postman\altaffilmark{1},
P. Rosati\altaffilmark{12}, 
M. Sirianni\altaffilmark{2}, 
W.B. Sparks\altaffilmark{1}, 
H.D. Tran\altaffilmark{13}, 
Z.I. Tsvetanov\altaffilmark{2},   
R.L. White\altaffilmark{1}
\& W. Zheng\altaffilmark{2}}

\altaffiltext{1}{Space Telescope Science Institute, 3700 San Martin Drive, Baltimore, MD 21218.}
\altaffiltext{2}{Department of Physics and Astronomy, Johns Hopkins
University, 3400 North Charles Street, Baltimore, MD 21218.}
\altaffiltext{3}{UCO/Lick Observatory, University of California, Santa
Cruz, CA 95064.}
\altaffiltext{4}{Bartko Science \& Technology, 14520 Akron Street, 
Brighton, CO 80602.}	
\altaffiltext{5}{Racah Institute of Physics, The Hebrew University,
Jerusalem, Israel 91904.}
\altaffiltext{6}{Conceptual Analytics, LLC, 8209 Woburn Abbey Road, Glenn Dale, MD 20769}
\altaffiltext{7}{NASA Goddard Space Flight Center, Code 681, Greenbelt, MD 20771.}
\altaffiltext{8}{Leiden Observatory, Postbus 9513, 2300 RA Leiden,
Netherlands.}
\altaffiltext{9}{Department of Astronomy and Astrophysics, The
Pennsylvania State University, 525 Davey Lab, University Park, PA
16802.}
\altaffiltext{10}{Departmento de Astronom\'{\i}a y Astrof\'{\i}sica,
Pontificia Universidad Cat\'{\o}lica de Chile, Casilla 306, Santiago
22, Chile.}
\altaffiltext{11}{Steward Observatory, University of Arizona, Tucson,
AZ 85721.}
\altaffiltext{12}{European Southern Observatory,
Karl-Schwarzschild-Strasse 2, D-85748 Garching, Germany.}
\altaffiltext{13}{W. M. Keck Observatory, 65-1120 Mamalahoa Hwy., 
Kamuela, HI 96743}

\begin{abstract}

We present {\it Hubble Space Telescope} Advanced Camera for Surveys multicolor
coronagraphic images of the recently discovered edge-on debris disk around the
nearby ($\sim10$ pc) M dwarf AU Microscopii.  The disk is seen between $r =
$0\farcs 75 -- 15'' (7.5 -- 150 AU) from the star.  It has a thin midplane
with a projected full-width-at-half-maximum (FWHM) thickness of 2.5 -- 3.5 AU
within $r < 50$ AU of the star that increases to 6.5 -- 9 AU at $r \sim 75$ AU.
The disk's radial brightness profile is generally flat for $r < 15$ AU, then
decreases gradually ($I \propto r^{-1.8}$) out to $r \approx 43$ AU, beyond
which it falls rapidly ($I \propto r^{-4.7}$).  Within 50 AU the midplane is
straight and aligned with the star, and beyond that it deviates by
$\sim3^{\circ}$, resulting in a bowed appearance that was also seen in
ground-based images.  Three-dimensional modelling of the disk shows that the
inner region ($r < 50$ AU) is inclined to the line-of-sight by $<1^{\circ}$ and
the outer disk by $\sim3^{\circ}$.  The inclination of the outer disk and
moderate forward scattering ($g \approx 0.4$) can explain the apparent bow.
The intrinsic, deprojected FWHM thickness is 1.5 -- 10 AU, increasing with
radius.  The models indicate that the disk is clear of dust within $\sim12$ AU
of the star, in general agreement with the previous prediction of 17 AU based
on the infrared spectral energy distribution.  The disk is blue, being 60\%
brighter at $B$ than $I$ relative to the star.  One possible explanation for
this is that there is a surplus of very small grains compared to other imaged
debris disks that have more neutral or red colors.  This may be due to the low
radiation pressure exerted by the late-type star.  Observations at two epochs
show that an extended source seen along the midplane is a background galaxy.  
 
\end{abstract}


\section{Introduction}

Collisions of planetesimals around a star produce dust grains, creating a
debris disk that can be detected in thermal emission and scattered light.
Without the continual replenishment of these grains, most of the disk will
disappear due to radiation and wind pressure, Poynting-Robertson drag, grain
coagulation, and tidal interactions with planets.  The lifetime of such disks
is not fully understood, and it may depend strongly on the luminosity of the
star.

While numerous detections and images of optically-thick accretion disks have
been made, only a few debris disks have been resolved.  The current inventory
of imaged debris disks is largely derived from {\it IRAS}-measured infrared
excesses of stars.  This list is biased toward stars of spectral types A -- F
because their luminosities are high enough to heat a debris disk to a {\it
IRAS}-detectable level.  More luminous stars would blow away such disks while
cooler stars cannot heat them enough to be detected.  Most of the resolved
debris disks surround A stars ($\beta$ Pictoris, Vega, Fomalhaut, HD 141569A,
and HR 4796), and only a few have been seen in scattered light.  The
sensitivity limits of {\it IRAS} leave uncertain the frequency of debris disks
around later-type stars.  The disk around $\epsilon$ Eridani, a K2V star with
one of the largest {\it IRAS}-measured stellar excesses, was detected by {\it
IRAS} only because of its very close proximity (3.2 pc) to us.  The {\it
Spitzer Space Telescope} should provide a much more comprehensive catalog of
disk candidates around later-type stars. 

There are only two M dwarfs, AU Microscopii and Hen 3-600 (Song et al. 2002),
not associated with molecular clouds that have {\it IRAS}-measured infrared
excesses.  AU Mic (HD 197481, GJ 803) is a M1Ve flare star and a BY Dra-type
variable ($V_{max} = 8.59, \Delta V \approx 0.15$; Cutispoto, Messina, \&
Rodon\`{o} 2003). Barrado y Navascu{\' e}s et al.  (1999) identified AU Mic as
part of the $\beta$ Pictoris moving group, an association of nearby (10 -- 50
pc) young stars.  It is at a {\it Hipparcos}-measured distance of 9.94 pc
(Perryman et al. 1997) and is about 10 Myr old (Zuckerman et al. 2001).  Its
excess implies that AU Mic has a substantial amount of circumstellar material.
These characteristics led the {\it Hubble Space Telescope} ({\it HST}) Advanced
Camera for Surveys (ACS) Investigation Definition Team (IDT) to include the
star in its circumstellar disk imaging program.  

A disk around AU Mic was recently imaged from the ground using coronagraphs
(Kalas, Liu, \& Matthews 2004, hereafter KLM04; Liu 2004, hereafter L04).
These images show a disk that at first glance resembles that of $\beta$ Pic.
It is viewed nearly edge-on and has a radius of at least 21'' (210 AU).  The
Keck adaptive optics image of L04 also shows small-scale variations in the
midplane that might be due to localized density enhancements caused by
unseen substellar companions that perturb the disk.  So far, imaging within
1\farcs 5 of the star has not been possible from the ground.  The high
resolution of ACS and its ability to see closer to the star can provide more
detailed views of this disk.

\section{Observations, Photometry, and Processing}

\subsection{Observations and Calibration}

The ACS IDT observations of AU Mic utilized the coronagraphic mode of the ACS
High Resolution Camera, which has a pixel scale of $\sim25$ mas pixel$^{-1}$
and a coronagraphic field point spread function (PSF)
full-width-at-half-maximum (FWHM) of 72 mas at $I$, 63 mas at $V$, and 50 mas
at $B$.  The $r = $ 0\farcs 9 occulting spot was used for all observations.
The coronagraph suppresses the diffraction pattern of the central star,
reducing the surface brightness of its wings by $\sim7\times$.  The remaining
flux from the occulted source is dominated by scattered light from the
telescope optics for $r > $ 1\farcs 5 and by light diffracted by the occulting
spot for $r < $ 1\farcs 5.  The latter appears as concentric rings around the
image of the spot.  Because of the stability of {\it HST}, most of the residual
light can be subtracted using an image of another star.
 
As AU Mic was on the ACS IDT target list before discovery of its disk, the
team's standard observation strategy for suspected disks was used: initial
imaging in one filter to confirm the existence of a disk, followed later with
multi-color imaging.  The ACS exposures are listed in Table 1.  The first
images ({\it HST} program 9987) were taken on 3 April 2004 in filter F606W
(wide $V$-band) and the second set (program 10330) on 24 July 2004 in F606W,
F435W (ACS $B$) and F814W (ACS wide $I$).  The telescope orientations were
specified to place the disk diagonally on the detector, and the orientations of
the two sets of images were separated by 90$^{\circ}$.  Exposures of HD 216149
($V$ = 5.4) in the same filters were taken after AU Mic at both epochs for use
in PSF subtraction.  This star was chosen because it is bright, is of a similar
color as AU Mic, and is near AU Mic on the sky so that any attitude-dependent
focus variations could be minimized.  In addition to the coronagraphic
exposures, short, direct (noncoronagraphic) images were taken to provide
photometry of the stars.

\begin{deluxetable}{lccrcc}
\footnotesize
\tablecaption{AU Mic and Reference PSF Exposures\label{tbl1}}
\tablewidth{0pt}
\tablehead{ \colhead{Star} & \colhead{Date} & \colhead{Filter} & 
\colhead{Exposure} & \colhead{Gain$^a$} & \colhead{Type$^b$} }
\startdata
   AU Mic  &  3 April 2004  &  F606W  &    2 x 0.1 s  & 4 & D \\
           &                &         &    3 x 750 s  & 2 & C \\
           &                &         &   2 x 1400 s  & 2 & C \\
           &  24 July 2004  &  F435W  &    1 x 0.1 s  & 4 & D \\
           &                &         &     1 x 60 s  & 2 & D \\
           &                &         & 4 x 1279.5 s  & 2 & C \\
           &                &  F606W  &    1 x 0.1 s  & 4 & D \\
           &                &         &     1 x 60 s  & 2 & D \\
           &                &         &    2 x 915 s  & 2 & C \\
           &                &  F814W  &    1 x 0.1 s  & 4 & D \\
           &                &         &     1 x 60 s  & 2 & D \\
           &                &         &    3 x 853 s  & 2 & C \\
           &                &         &               &   &   \\
HD 216149  &  3 April 2004  &  F606W  &    2 x 0.1 s  & 4 & D \\
           &                &         &     2 x 90 s  & 2 & C \\
           &                &         &    8 x 225 s  & 2 & C \\
           &  24 July 2004  &  F435W  &    1 x 0.1 s  & 4 & D \\
           &                &         &      2 x 5 s  & 2 & D \\
           &                &         &    2 x 475 s  & 2 & C \\
           &                &  F606W  &    1 x 0.1 s  & 4 & D \\
           &                &         &      2 x 5 s  & 2 & D \\
           &                &         &    2 x 145 s  & 2 & C \\
           &                &  F814W  &    1 x 0.1 s  & 4 & D \\
           &                &         &      2 x 5 s  & 2 & D \\
           &                &         &    2 x 145 s  & 2 & C
\enddata
\tablenotetext{a}{In electrons per data number}
\tablenotetext{b}{D = direct (noncoronagraphic) and C = coronagraphic}
\end{deluxetable}

All of the images were calibrated by the {\it HST} pipeline.  To account for
vignetting around the occulting spot, the coronagraphic images were manually
divided by the spot flats available from the ACS reference files page at the
Space Telescope Science Institute (STScI) web site.  Because the spot varies in
position over time, each spot flat was shifted to the appropriate position as
listed on the ACS web page.  The duplicate exposures were combined with cosmic
ray rejection.  At this stage, the images were not corrected for geometric
distortion.

\subsection{Stellar Photometry}

Photometric measurements of the two stars are required to properly normalize
the HD 216149 images prior to subtraction from those of AU Mic.  The AU Mic
fluxes were measured from the short, noncoronagraphic images using circular
apertures with radii of 13 (F435W), 25 (F606W), and 30 (F814W) pixels,
increasing in size to account for the larger PSFs at longer wavelengths.  The
short exposures of HD 216149 were saturated.  As reported by Gilliland (2004)
\footnote{ACS Instrument Science Report ISR 04-01 (Gilliland 2004) is available
at http://www.stsci.edu/hst/acs/documents/isrs.}, the well depth of an HRC
pixel is $\sim$160,000 e$^-$, which is within the 16-bit range of the
electronics when a gain of 4 e$^-$ per data unit is used.  Saturating the pixel
does not saturate the analog-to-digital converter, so flux is not lost when a
pixel saturates but instead overflows into an adjacent pixel.  As Gilliland
(2004) showed, in this case the counts are conserved, and summation within an
aperture that has a radius that encompasses all of the saturated pixels should
provide the same countrate as in an unsaturated image.  The fluxes for HD
216149 in the three filters were measured using 30 pixel radius apertures,
which include all of the saturated pixels.

The measured photometry for both stars were corrected to effectively infinite
apertures using conversion coefficients derived from encircled energy curves of
well-exposed, high-dynamic-range images of other stars.  Because the AU Mic and
HD 216149 measurements were made using images that had not been corrected for
ACS geometrical distortion, compensation for pixel-area-distortion was made
using the pixel area maps provided by STScI (Pavlovsky et al. 2004)\footnote{
ACS Data Handbook (Pavlovsky et al. 2004) is available at
http://www.stsci.edu/acs/documents.}.  The instrumental fluxes were converted
to standard {\it BVI}$_c$ magnitudes (Table 2) using the STSDAS SYNPHOT
synthetic photometry program assuming M1V (for AU Mic) and M0III (for HD
216149) spectra. 

\begin{deluxetable}{lrcc}
\footnotesize
\tablecaption{AU Mic and Reference PSF Photometry\label{tbl2}}
\tablewidth{0pt}
\tablehead{ \colhead{Star} & \colhead{Date} & \colhead{Passband} & 
\colhead{Magnitude$^a$} }
\startdata
AU Mic    &  3 April 2004  &  $V$    &  8.63 $\pm$0.03 \\
          &  24 July 2004  &  $B$    &  9.96 $\pm$0.05 \\
          &                &  $V$    &  8.64 $\pm$0.03 \\
          &                &  $I_c$  &  6.60 $\pm$0.03 \\
          &                &         &                 \\
HD 216149 &  3 April 2004  &  $V$    &  5.46 $\pm$0.03 \\
          &  24 July 2004  &  $B$    &  6.71 $\pm$0.05 \\
          &                &  $V$    &  5.46 $\pm$0.03 \\
          &                &  $I_c$  &  4.05 $\pm$0.03   
\enddata
\tablenotetext{a}{Stated error includes estimate of magnitude system 
transformation error}
\end{deluxetable}

\subsection{PSF Subtraction}

The HD 216149 coronagraphic images were normalized to match the measured AU Mic
fluxes.  They were then iteratively shifted by subpixel amounts via cubic
convolution interpolation and subtracted from the AU Mic images until the
residual scattered light patterns were visibly minimized.  This procedure
appears to align the two stars to within $\pm$0.05 pix, aided by the
high-spatial-frequency streaks in the residual pattern, which can be seen
radiating from the star in Figures 1c-1f.  After subtraction, the images were
corrected for geometric distortion.  The diffraction patterns inside and around
the edge of the occulting spot were not well subtracted.  Their shapes and
intensities are quite sensitive to the alignment of the star behind the spot,
which at worst differed by 16 mas between AU Mic and the reference PSF.  They
are also sensitive to focus changes and small differences between the spectra
of the stars over a filter's bandpass.  Their residuals, which appear as
alternating positive and negative rings, are the dominant source of error for
$r < $ 1\farcs 5, where they create localized uncertainties of $\pm30\%$ on
0\farcs 1 scales in the disk brightness in the F606W images.  While there are
oscillations within and around the spot in both F606W images, they appear to
have mean residuals near zero.  However, in F814W the spot interior appears
oversubtracted, and in F435W it is undersubtracted, probably due to focus
mismatches.  Further out the background is dominated by scatter-subtraction
residuals, which cause pixel-to-pixel errors of $<10\%$.  

In addition to the coronagraphic images, we also subtracted the longest
noncoronagraphic exposure of HD 216149 from that of AU Mic.  While this clearly
revealed the disk, it did not allow detection of it within 1'' of the star due
to high subtraction residuals, providing no improvement over the coronagraphic
subtractions.  We also used the two F606W coronagraphic images of AU Mic to
subtract each other, but due to differences in the spot positions between the
two epochs, the subtraction residuals were larger than those when HD 216149 was
used.

\begin{figure} 
\includegraphics[width=6in]{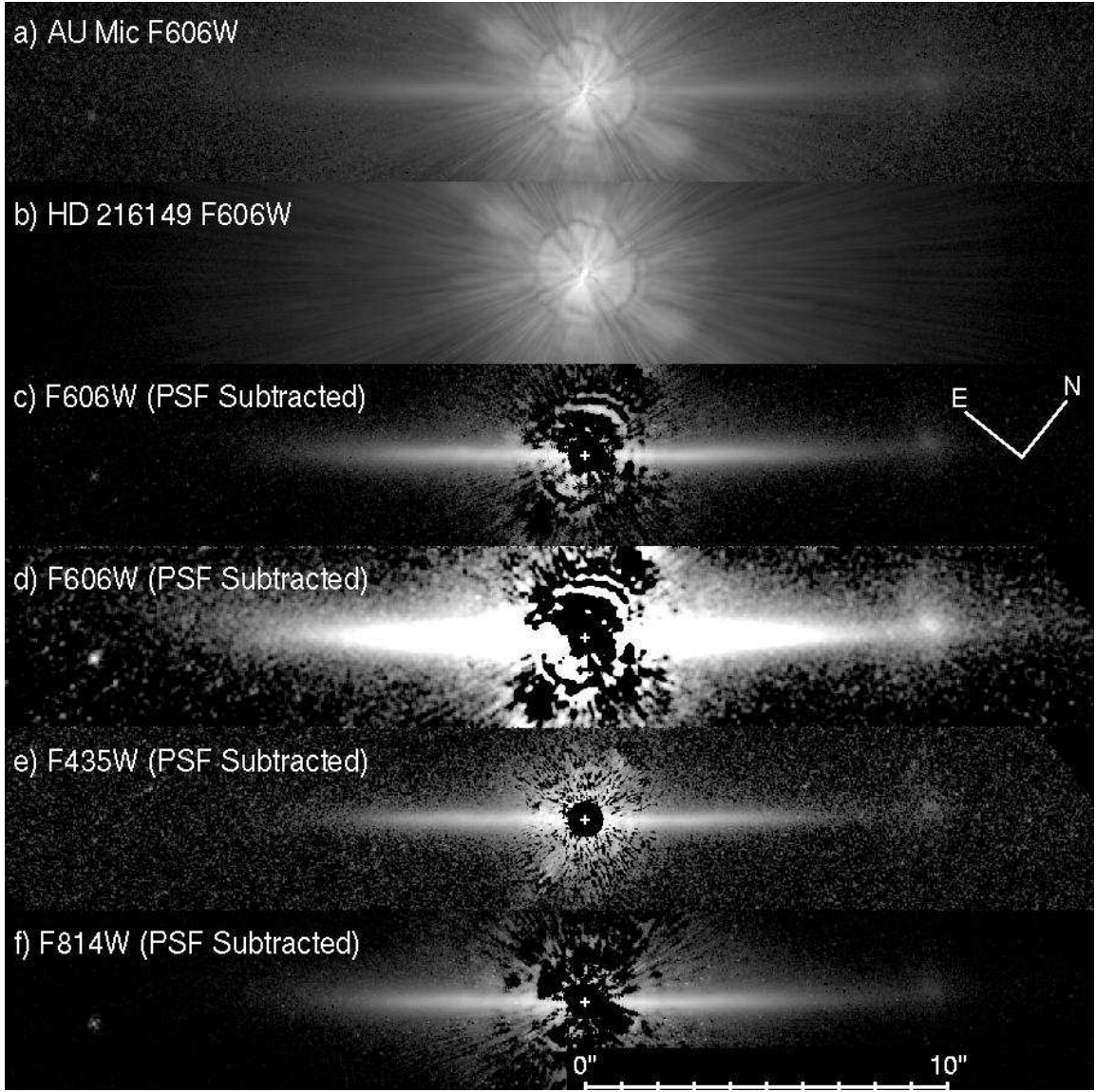} 
\caption{First epoch ACS F606W coronagraphic images of AU Mic and the reference
PSF star HD 216149, all displayed with logarithmic intensity scaling. Image (d) 
is smoothed and shown using a stronger stretch to reveal the disk at greater
heights and structure associated with the galaxy superposed on the NW
extension.  A background star can be seen below the SE extension.} 
\label{fig1}
\end{figure}

\section{Results}

The AU Mic disk stands out against the background PSF structure in the ACS
coronagraphic images (Figure 1a) even before PSF subtraction.  In the first
epoch F606W image (the deepest exposure) the northwestern side is detected out
to the edge of the detector, 15\farcs 25 ($\sim$150 AU) from the star, and the
southeastern side to $\sim$14'' ($\sim$140 AU).  The disk can be seen to a
height of 2\farcs 5 (25 AU) above the midplane.  The observed radial and height
extents in this image are sensitivity-limited.  The interior of the occulting
spot is filled with light that was not blocked by the spot (which is located in
the aberrated {\it HST} beam) and was afterward modified by the corrective ACS
optics.  An image of the star at the spot's center allows accurate measurement
of the stellar position.  An extended source, which shall later be shown to be
a background galaxy, can be seen superposed on the NW midplane, 9\farcs 6 from
the star. A point source is also seen $\sim1$'' SW of the disk midplane
13\farcs7 SE from the star.  A few background galaxies are also seen throughout
the field. 

The PSF-subtracted image more clearly reveals the disk (Figure 1c-f).  Because
the disk is fairly bright close to the star and the corrective optics modify
the aberrated light that passes by the occulter, the disk can actually be seen
inside the spot, to within $\sim$0\farcs 75 of the star.  The residual levels
indicate that using the coronagraph with PSF subtraction improves the
disk-to-background contrast by $\sim150\times$ compared to direct imaging
without subtraction.  Significant residuals are present in a bar of
instrumentally-scattered light that extends from the upper left to lower right
of the coronagraphic field, but these are positioned away from the disk.  The
F606W subtractions from each epoch are of comparable quality, while the F814W
subtraction has large residuals near and within the spot, probably due to focus
differences between the two images.  The residuals in the F435W image are more
uniform than in the other two filters, but because AU Mic is much fainter in
that passband, the effects of electronic noise are more significant. 

The extended source detection limits in the subtracted images were estimated by
adding 1'' $\times$ 1'' uniform-intensity squares to the data at 3'' and 9''
from the star perpendicular to the circumstellar disk.  The squares'
intensities were adjusted until they could no longer be visually detected.  The
derived visual detection limits are $B$ = 23.3 and 23.5 mag arcsec$^{-2}$
(F435W), $V$ = 22.6 and 24.0 (F606W), and $I$ = 20.4 and 21.9 (F814W), for $r =
$ 3'' and 9'', respectively, with estimated errors of $\pm0.2$ mag
arcsec$^{-2}$.

\begin{figure} 
\includegraphics[width=6in]{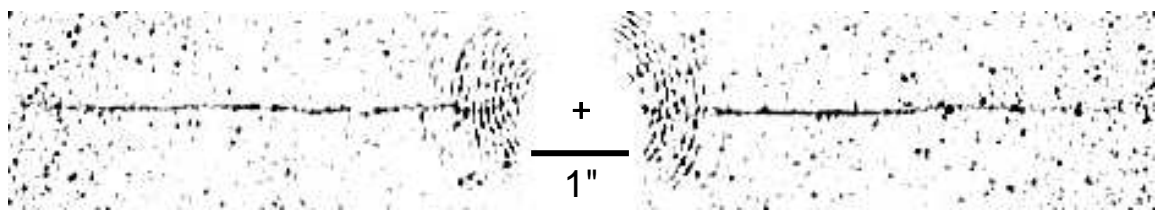}
\caption{First epoch ACS F606W coronagraphic PSF-subtracted image of AU Mic
after subtraction of a median-smoothed copy of itself, highlighting the higher
spatial frequency structures, notably the flat, sharp disk midplane (other
features are noise and PSF subtraction residuals).  Inverse intensity and the
same orientation as Figure 1.} 
\label{fig2}
\end{figure}

\subsection{Disk Morphology}

Analysis of the disk morphology is largely confined to the first epoch F606W
observation, as it was the first and deepest exposure.  As will be shown in
Section 3.3, there are no significant differences in the disk morphology among
the three filters.

To allow accurate measurement of the disk orientation, simple spatial filtering
was applied to highlight the midplane.  The image was subtracted by a copy of
itself that was smoothed by a $3 \times 3$ pixel median filter.  This removed
the extended vertical wings while preserving the sharp peak of the midplane
seen within 5'' of the star (Figure 2).  A line fit to the resulting image
shows that the inner ($r < 5$'') midplane is oriented at PA=128.6$^{\circ} \pm
0.2^{\circ}$.  There are small ($\sim40$ mas), oscillations in the midplane
position about this line, which passes directly through the star.  Superposing
this line over the original image highlights the deviation of the outer ($r >
5$'') midplane from the inner (Figure 3).  At large angles, the disk looks
``bowed'', an effect noted in the ground-based images and also seen in the
$\beta$ Pic disk (Kalas \& Jewitt 1995).  The apparent outer disk midplane
measured by-eye at $r$=12\farcs 5 is oriented along PA=$311.1^{\circ} \pm
0.7^{\circ}$ on the NW side and PA=$125.6^{\circ} \pm 0.7^{\circ}$ on the SE.
These angles are consistent with those reported by KLM04 and L04.

\begin{figure} 
\includegraphics[width=6in]{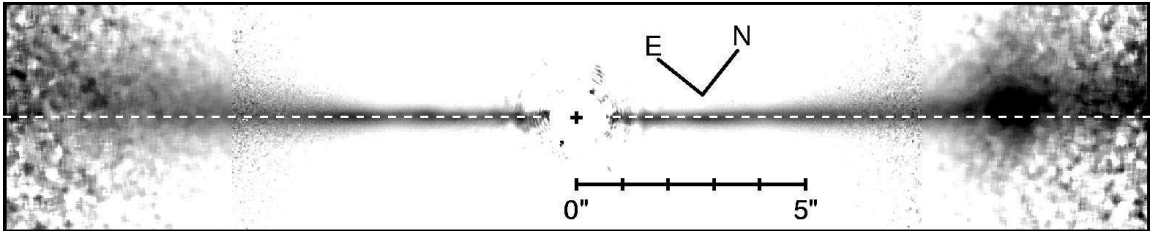} 
\caption{First epoch ACS F606W coronagraphic image of the AU Mic disk
normalized by the radial brightness profile power laws described in the paper
and displayed in inverse intensity.  Data beyond 7\farcs 5 has been
median-smoothed.  The ``bowing'' can be seen in the outer disk as an upward
deviation from the horizontal line defined by the inner midplane.} 
\label{fig3}
\end{figure}

\begin{figure} 
\includegraphics{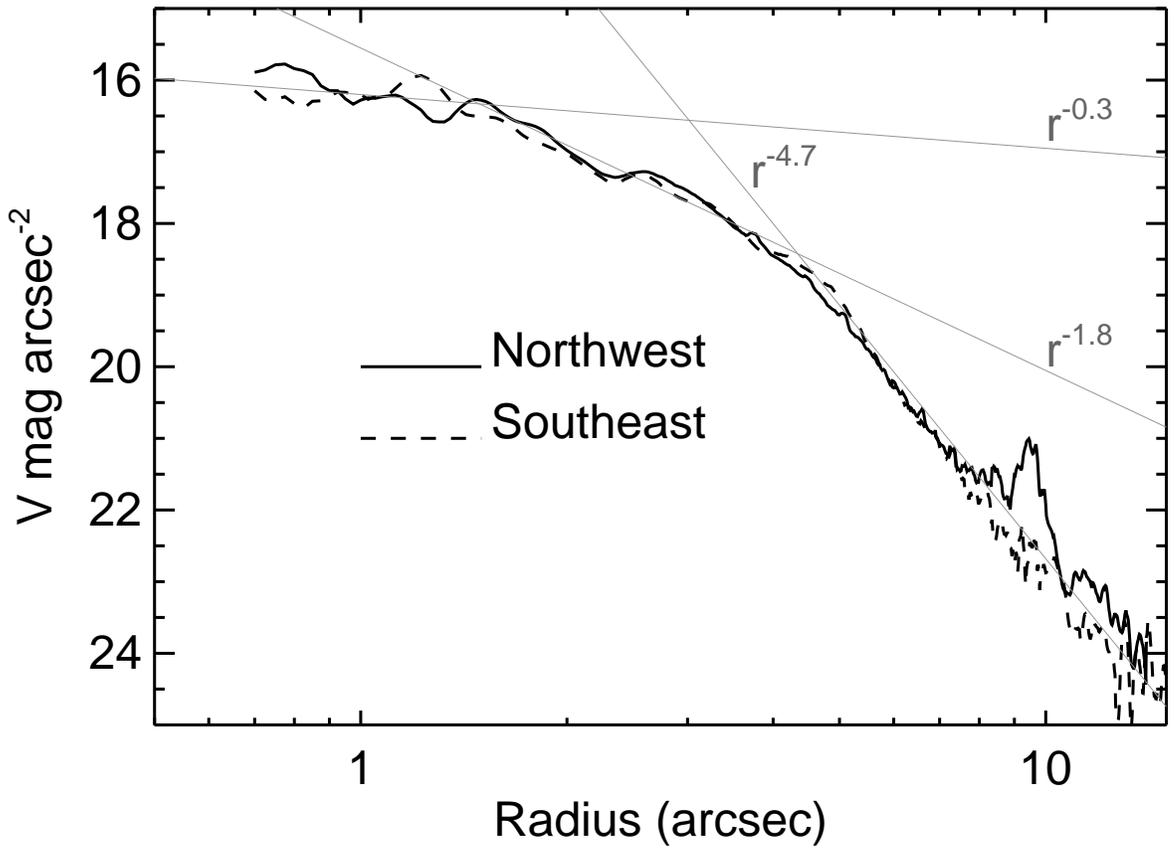}
\caption{Radial surface brightness profiles following the AU Mic disk midplane
(including the bow) measured over 0\farcs 25 $\times$ 0\farcs 25 pixels in the
first epoch ACS F606W coronagraphic image.  Power law profiles are
overplotted.} 
\label{fig4}
\end{figure}

Radial surface brightness profiles along the disk midplane, accounting for the
bow, are shown in Figure 4.  Both sides of the disk appear to have very
similar profiles, though the NW side is brighter beyond 10'' (100 AU) by
$\sim2\times$ relative to the SE.  The mean midplane radial brightness profile
may be reasonably divided into three zones described by different power laws.
The inner zone ($r < $ 1\farcs 5) is nearly flat ($I \propto r^{-0.3}$).  The
middle zone (1\farcs 5 $<  r < $ 4\farcs 3) has a moderate radial decrease in
brightness ($I \propto r^{-1.8}$), which is slightly steeper than that measured
by L04 ($r^{-1.0}$ to $r^{-1.4}$).  The outer zone ($r >$ 4\farcs 3) drops off
rapidly ($I \propto r^{-4.7}$), which is steeper than the $\sim r^{-3.8}$
measured by KLM04 between 6'' -- 16'' but consistent with the $r^{-4.4\pm0.3}$
reported by L04. 

Localized deviations from the large-scale brightness profile are present.
There are $\sim20\%$ dips at $r$ = 2\farcs 3 (23 AU) on both sides that appear
to be genuine and not PSF subtraction artifacts.  Between 4\farcs 3 and 5\farcs
3 the SE side is $\sim25\%$ brighter than the NW.  Within 1\farcs 5 the
subtraction residuals are too large to identify any localized asymmetries of
less than 30\%.  To highlight the small-scale variations, each row in the image
parallel to the midplane was divided by a smooth function, a 4$^{th}$-order
polynomial fit derived from the mean midplane radial brightness profile.  The
resulting image (Figure 5) emphasizes the deviations and can be compared to
Figure 3 of L04.  The dip at 23 AU is clear, though any features closer to the
star than it are suspect.  There are correlations between features in these
images and those seen in the L04 data (L04 labeling convention used; distances
in parentheses are the centers of the features as seen in the L04 image): (A) a
``clump'' at 26 AU (25 AU), (B) a dip at 29 AU (29 AU), and (C) a clump at 33
AU (31 AU).  The relatively large discrepancy in the position of feature (C)
between the {\it HST} and L04 data may be due to subtraction errors in either
or both datasets.  We note that the positions of the features appear the same
in the {\it HST} images at both epochs.  At 37 AU on the NW side there is a
marginal enhancement that may correspond to L04's clump D, while at the same
radius in the SE side there is a slight dip.  Beyond that radius the NW side
smoothly declines, but there is one more extended clump at 48 AU in the SE.

\begin{figure} 
\includegraphics[width=6in]{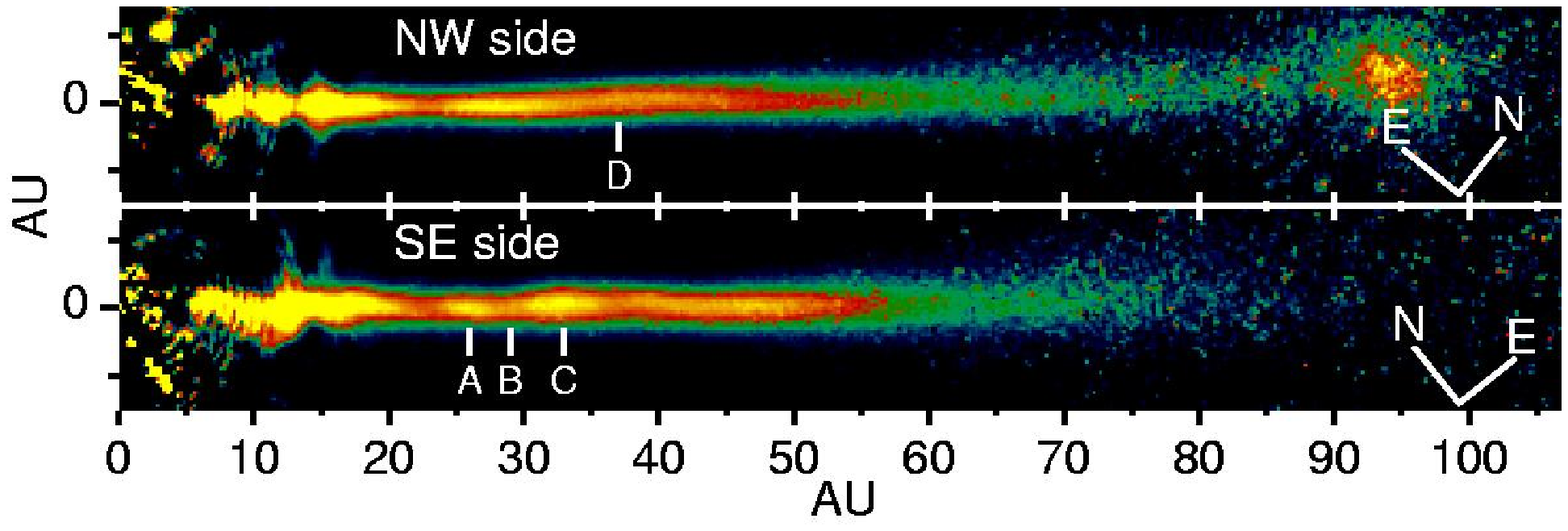}
\caption{AU Mic disk ACS F606W image divided by a smooth function fit to the
midplane to highlight localized variations.  The SE side has been flipped 
about the star for easier comparison.}
\label{fig5}
\end{figure}

\begin{figure} 
\includegraphics{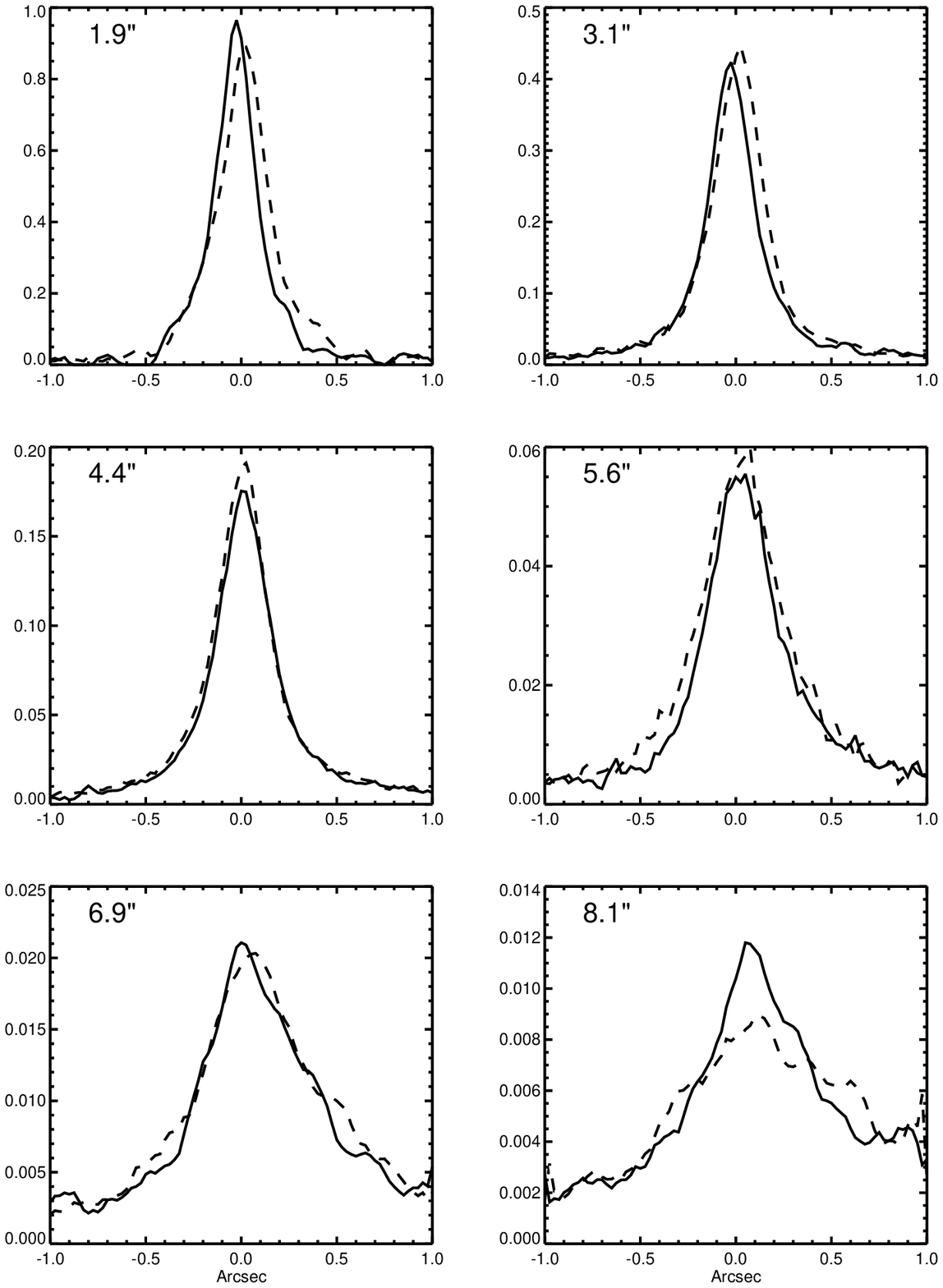}
\caption{AU Mic disk intensity cross-sections perpendicular to the midplane at
various radii from the first epoch ACS F606W coronagraphic image.  The solid
lines represent the NW extension of the disk and dashed the SE.  The NE side 
of the disk is toward the right.  The disk bow can be seen as the shift in
flux toward the NE at larger radii.} 
\label{fig6}
\end{figure}

The inner disk's vertical brightness profile (Figure 6) shows a sharp midplane
with extended wings.  It can be reasonably well characterized for disk radii
$<$5'' by a Lorentzian profile: $I(z) = 1 / (1 + z^2/h^2)$, where $2h$ is the
profile full-width-at-half-maximum (FWHM) and $z$ the height above the
midplane.  Note that we find this particular shape to be a convenient
description of the disk profile and do not assert any physical significance to
it.  The profiles are generally symmetric about the midplane for $r<5''$ but
beyond this the disk is brighter on the NE side of the midplane.  The outer
disk profile is more rounded, lacking a sharp midplane.

\begin{figure} 
\includegraphics{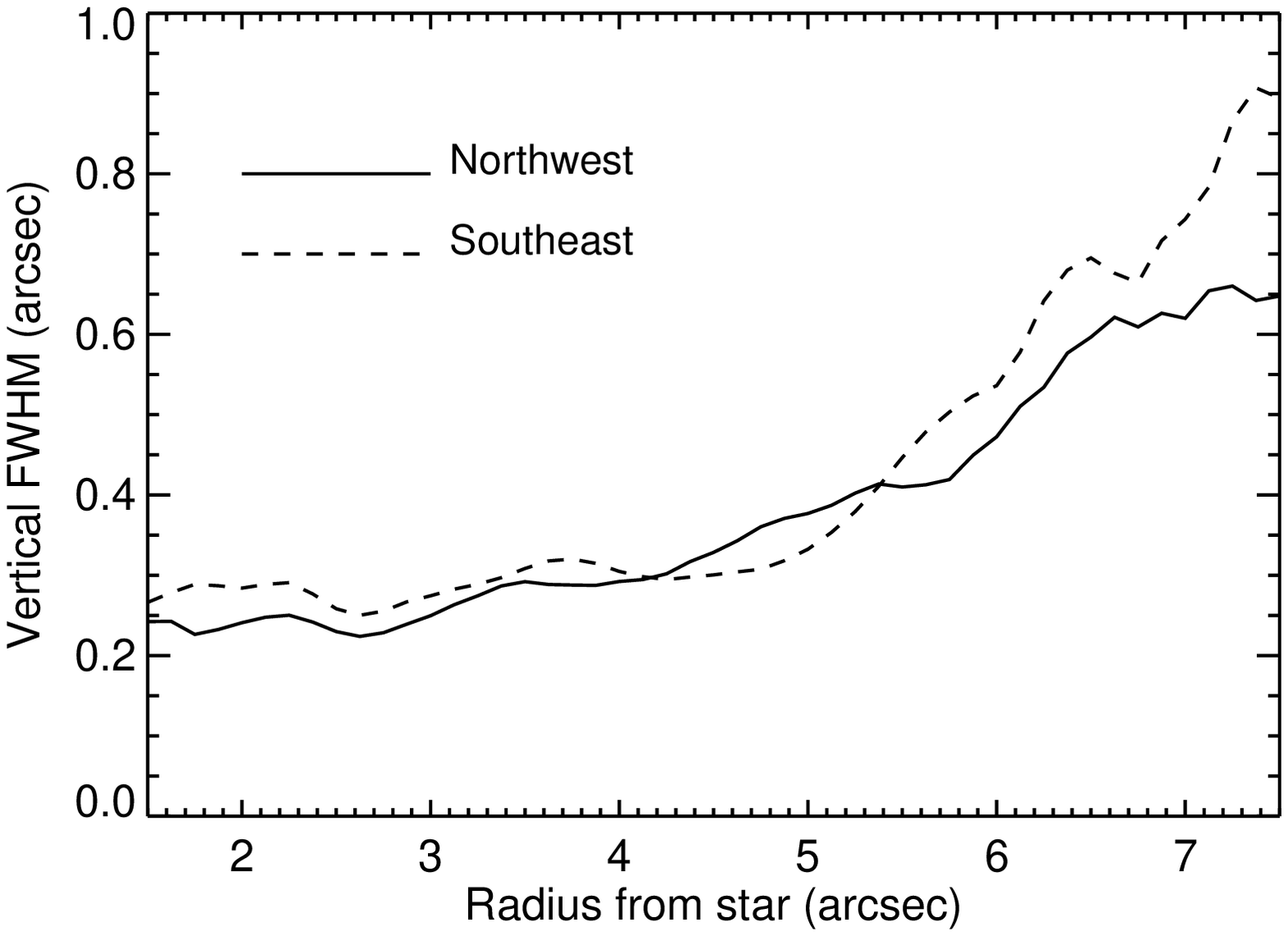} 
\caption{AU Mic disk FWHM measured from the ACS F606W coronagraphic image.} 
\label{fig7}
\end{figure}

The FWHM at each radius was measured by fitting a Lorentzian profile to a
vertical cross-section of the image after smoothing by 5 pixels along the
direction of the midplane.  As shown in Figure 7, the inner disk is rather
flat, but the outer disk begins to rapidly broaden beyond 5''.  The FWHM for $r
< 5''$ (50 AU) is $0\farcs 25 - 0\farcs 35$ ($2.5 - 3.5$ AU).  At $r = 7\farcs
5$ (75 AU) the FWHM on the SE side is 0\farcs 9 (9 AU) and 0\farcs 65 (6.5 AU)
on the NW.  The change in FWHM along the SE side can be approximately described
by two power laws: $FWHM \propto r^{0.07}$ for $1\farcs 5 < r < 4\farcs 6$ and
$r^{2.4}$ for $4\farcs 6 < r < 7\farcs 5$.  The equivalent relations for the NW
side are $FWHM \propto r^{0.07}$ for 1\farcs 5 -- 3\farcs 0 and $r^{1.0}$ for
3\farcs 0 -- 7\farcs 5.  These apparent vertical profiles represent the
projection of the optically-thin disk along the line of sight convolved with
the instrumental PSF.

\subsection{Disk Modelling}

Derivation of the physical distribution and properties of the dust in the disk
directly from an image is prohibited by the integrated effects of forward
scattering and density variations seen along the line of sight.  They can be
estimated from three-dimensional scattering models optimized to match the
observed image.  We have attempted to derive a reasonable set of model
parameters that match the major features of the disk in the first-epoch F606W
exposure.  We emphasize that we have not fully explored the parameter space,
which is likely to have some degeneracies that prevent determination of a
unique set of optimal values. 

The modelling code computes the singly-scattered reflected light from the disk
integrated along the line of sight.  Because the AU Mic disk is optically thin
(KLM04), nearly all of the observed light is singly-scattered, and the fraction
of multiply-scattered photons should be insignificant.  Assuming axial
symmetry, the scattering surface density $\rho(r,z)$ at radius $r$ and height
$z$ is characterized by $\rho(r,z) = \rho_0 (r/r_0)^{\alpha} \phi(z,h)$ where
$\rho_0$ is the midplane scattering density at some fiducial radius $r_0$ and
the scale height of the vertical distribution profile $\phi(z)$ varies as $h =
h_0 (r / r_0)^\beta$.  An albedo of 0.5 is assumed.  We used a Lorentzian
vertical density distribution profile where $\phi(z) = 1 / (1 + (z/h)^2)$.  The
code can compute a model with contiguous annular zones, each with independent
$\alpha$, $\beta$, inclination, midplane position angle, and outer radius (the
outer radius of one zone defines the inner radius of the next).  The midplane
densities and scale heights are constrained to be equal at the interface
between two zones.  The inner radius of the innermost zone is also defined,
allowing for a clear central region.  The intensity of reflected light from
each point in the three-dimensional model is modified by the Henyey-Greenstein
scattering phase function.  The scattering asymmetry parameter $g$ defines how
forward-scattering the grains are ($g$ = 0 is isotropic, $g$ = 1 is fully
forward-scattering). 

An iterative, non-linear least squares fitting routine was used to optimize the
disk parameters by generating models and comparing them to the image.  Both
sides of the disk were simultaneously fit using the model, which is symmetric
about the star within each disk zone.  To avoid the brighter inner regions from
completely driving the results, the weights for all of the pixels at each
radius were multiplied by the inverse of the peak midplane intensity at that
radius.  Regions covered by the galaxy at $r = 9\farcs6$ and within 0\farcs 75
of AU Mic were given zero weight.  The disk model was convolved with a Tiny Tim
PSF model (Krist \& Hook 2004)\footnote{The Tiny Tim software and manual are
available from http://www.stsci.edu/software/tinytim.} to account for
instrumental blur.  During the comparisons pixels at $r > 7\farcs 5$ in both
the data and model were smoothed using a $7 \times 7$ pixel median filter to
reduce the noise in those faint regions. The apparent disk midplane is
sufficiently extended at those radii to be unaffected by this smoothing.  All
of these models were computed with a 250 AU outer radius and 25 AU maximum
height.

The model disk was initially composed of three annular zones (1 -- 3)
surrounding an inner clearing.  The inner radius was somewhat arbitrarily set
at 3 AU, within the masked central region.  The outer radii (which were free
parameters) of the zones 1 -- 3 were set to 14, 45, and 250 AU, respectively,
corresponding to the regions defined by the power law descriptions of the
radial brightness profile.  The starting values of $\alpha$ were 0.0 for the
two inner zones and -4.0 for the outer.  Likewise, $\beta$ was started at 0.0
for the two inner zones and 2.0 for the outer.  $h_0$ and $\rho_0$ were
determined by trial-and-error adjustments until a reasonable initial model was
produced.  Separate fits were made with $g$ starting at 0.0, 0.3, and 0.6.  In
these runs the zones were constrained to have the same inclination (starting at
0.0) and position angle.  We refer to these simple three-zone models
collectively as Model 1. 

The first runs indicated that Zone 1 was essentially clear, as the values for
$\alpha$ in all of the fits were extremely large, corresponding to a very rapid
density increase at the outer boundary of the zone.  The values of $g$ were
0.45 -- 0.65, and the inclination offsets from edge-on were 0.0$^{\circ}$ --
0.8$^{\circ}$.  The values for $\beta$ varied chaotically for the inner two
zones, and the $(data - model)$ residuals showed that the localized variations
were likely causing difficulties in determining $\beta$ there.  In the outer
zone $\beta$ = 2.1 -- 2.6.  None of the resulting models reproduced the bow or
the change in the vertical profile shape between the inner and outer disk, so
these fits were not deemed acceptable.

A new fit was made with some modifications to the zone layout.  First, Zone 1
was removed and replaced with an expanded inner clearing with an initial outer
radius of 14 AU.  A new zone with an initial outer radius of 85 AU was inserted
between the outer two zones in order to provide more freedom to fit the
transition region between the inner and outer disk.  Finally, each zone was
allowed to have its own inclination and position angle.  We refer to this more
complicated model as Model 2.  As before, separate fits were run with $g$
starting at 0.0, 0.3, and 0.6 and with the same initial parameters as Model 1.  

While they cannot match the small-scale variations in the disk (Figure 8), the
new fits provide improved models that reproduce the bow (Figure 9), the
radial brightness profile (Figure 10), and the change in the vertical profile
shape with radius (Figure 11).  The parameter ranges are given in Table 3.
Despite starting with significantly different parameters, the separate fits
returned very similar results.  The inner clearing is $\sim12$ AU in radius,
which is similar to the $\sim17$ AU predicted by fitting the spectral energy
distribution (Liu et al. 2004).  The inner annular zone (Zone 1) extends to
$\sim49$ AU and has both a flat radial density distribution and thickness
($\alpha$ and $\beta$ are both approximately zero).  It is quite close to
edge-on ($i \approx 0.6^{\circ}$), as is Zone 2 ($i \approx -0.5^{\circ}$).
The density falls rapidly with radius over Zone 2 ($\alpha \approx -4.6$) while
the disk thickness increases ($\beta \approx 2.5$).  The density falls off less
rapidly over the outermost zone ($\alpha \approx -2.6$) and the thickness
increases slowly ($\beta = 0 \sim 0.6$), but the inclination is notably greater
with $i \approx -3^{\circ}$ (the SW pole is inclined toward us).  The position
angles of the zones are all within 0.5$^{\circ}$ of each other. 

The model shows that forward scattering ($g \approx 0.4$) and the inclination
of the outer disk ($r > 80$ AU) cause the bow.  We do not believe that any
other processes, such as interaction with the ISM (Liu 2004) are required to
explain its appearance.  The disk is intrinsically quite thin, with a FWHM
thickness of $\sim1.8$ AU at $r = 20$ AU, compared to a projected and
PSF-convolved FWHM of $\sim2.5$ AU.  It broadens to $\sim6$ AU at at $r = 80$
AU.  The model suggests that the outer disk may have the same vertical profile
as the inner, but due to its higher inclination it appears to lack a sharp
midplane. 
 
\begin{deluxetable}{lcc}
\footnotesize
\tablecaption{AU Mic Disk Model Fit Parameters\label{tbl3}}
\tablewidth{0pt}
\tablehead{ \colhead{Parameter} & \colhead{Minimum} & \colhead{Maximum} }
\startdata
Midplane scattering cross-section     &     &  \\
\ \ \ density ($r = 20$ AU) (cm$^{-2}$/cm$^{-3}$)  & $1.117 \times 10^{-15}$  & $1.120 \times 10^{-15}$   \\  
FWHM thickness (AU) at $r = 20$ AU  & 1.73 & 1.74 \\
Scattering asymmetry ($g$)  &  0.372  & 0.374 \\
Inner disk radius (AU)      &  11.70  & 11.71 \\
                            &         &       \\
Zone 1:                &         &       \\
Outer radius (AU)           &  48.6   & 49.1  \\
Offset from edge-on         &  0.56$^{\circ}$  & 0.57$^{\circ}$ \\
$\alpha$                    & -0.066  &  -0.053 \\
$\beta$                     & 0.074   &  0.085  \\
                            &         &       \\
Zone 2:               &         &       \\
Outer radius (AU)           &  78.9   & 82.4  \\
Offset from edge-on         &  -0.52$^{\circ}$  & -0.45$^{\circ}$ \\
$\alpha$                    & -4.66   &  -4.58 \\
$\beta$                     & 2.45    &  2.52  \\
                            &         &       \\
Zone 3:               &         &       \\
Outer radius (AU)           &  250    & 250   \\
Offset from edge-on         &  -2.97$^{\circ}$  & -2.93$^{\circ}$ \\
$\alpha$                    & -2.68   &  -2.31 \\
$\beta$                     & 0.01    &  0.59  \\
\enddata
\end{deluxetable}

\begin{figure} 
\includegraphics[width=6in]{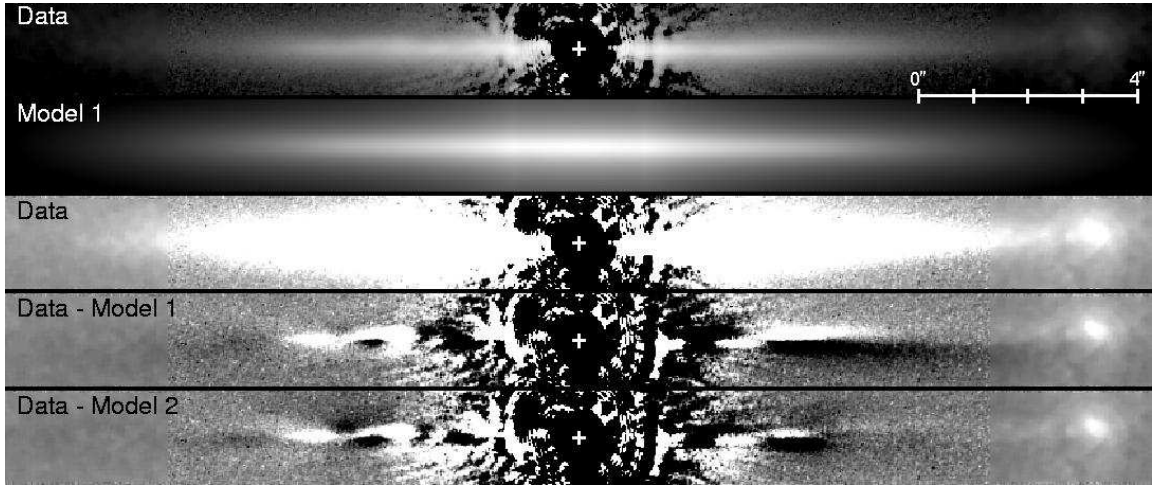}
\caption{Comparisons of the AU Mic disk ACS F606W image with three-dimensional
scattering models.  The top two images show the observed data and the simple,
three-zone model (Model 1) displayed with the same logarithmic intensity
stretch.  Below them is the observed image and the (Data - Model) difference
images displayed with the same strong linear stretch.  Model 2 is the more
complex model in which each zone may have different inclinations.  The cross
marks the position of the star.  Data beyond 7\farcs 5 is median-smoothed.
Same orientation as Figure 1.} 
\label{fig8}
\end{figure}

\begin{figure} 
\includegraphics[width=6in]{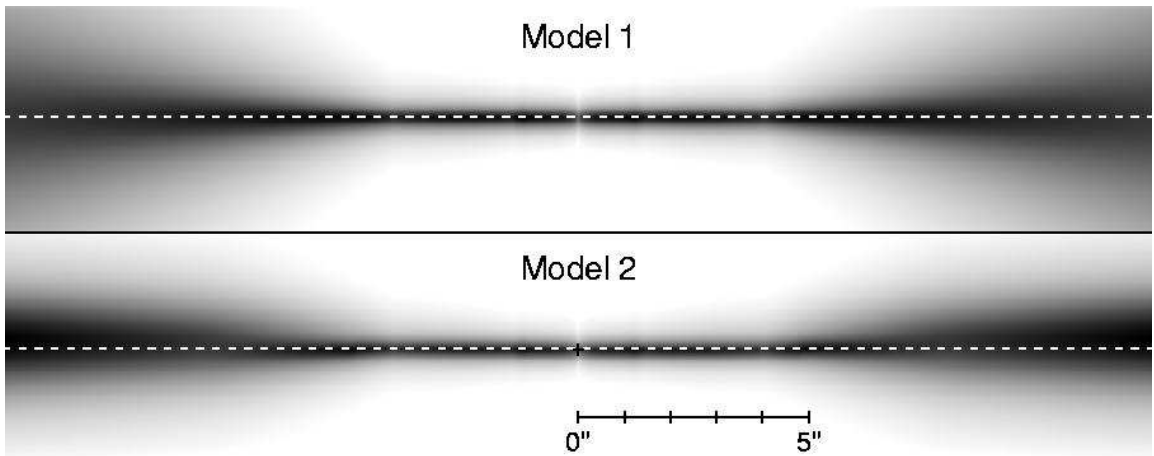} 
\caption{Model 1 (coplanar zones) and Model 2 (independently-inclined zones)
displayed with inverse intensities showing how Model 2 better reproduces the
disk bow (the outer disk midplane appears offset toward the top).  This can
be compared to the observed bow shown in Figure 3 (same orientation as Figure
3).}
\label{fig9}
\end{figure}

\begin{figure} 
\includegraphics{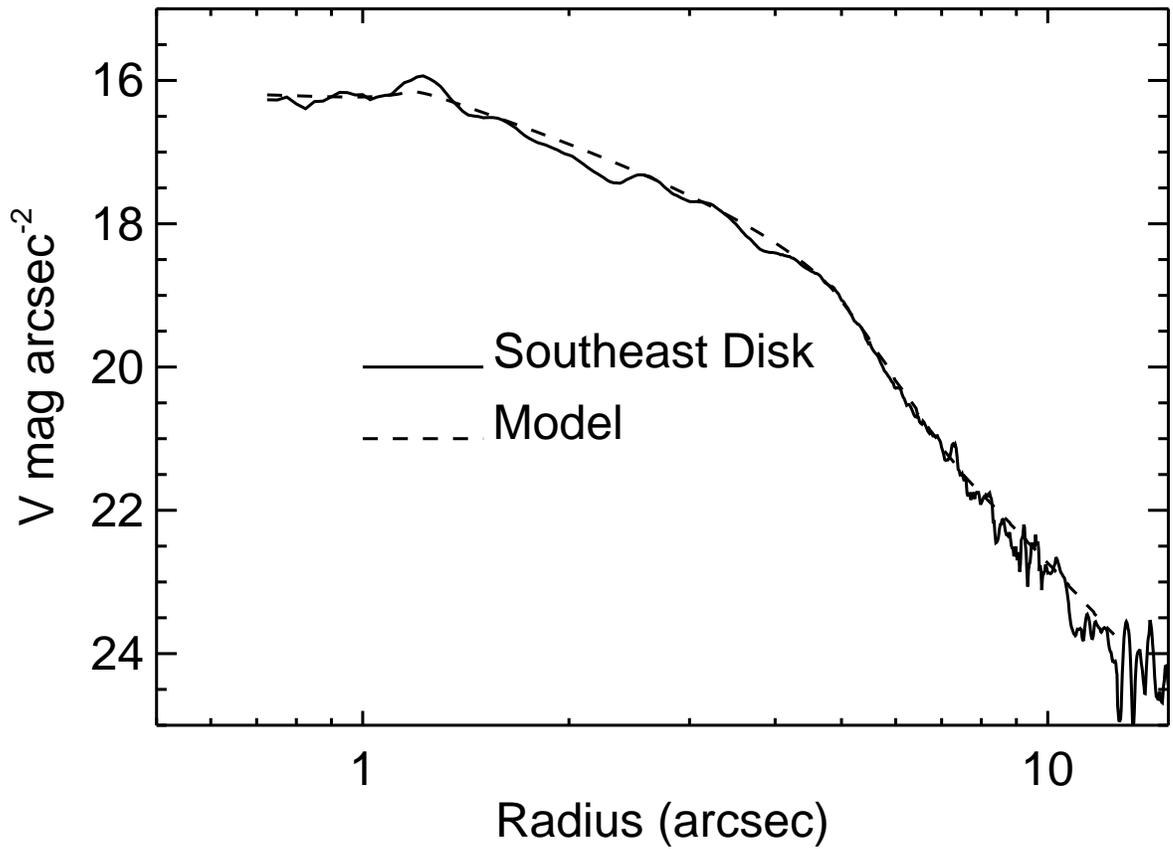} 
\caption{Comparison of the ACS F606W radial surface brightness profiles of the
SE side of the disk and the equivalent section of the final model.} 
\label{fig10}
\end{figure}

\begin{figure} 
\includegraphics{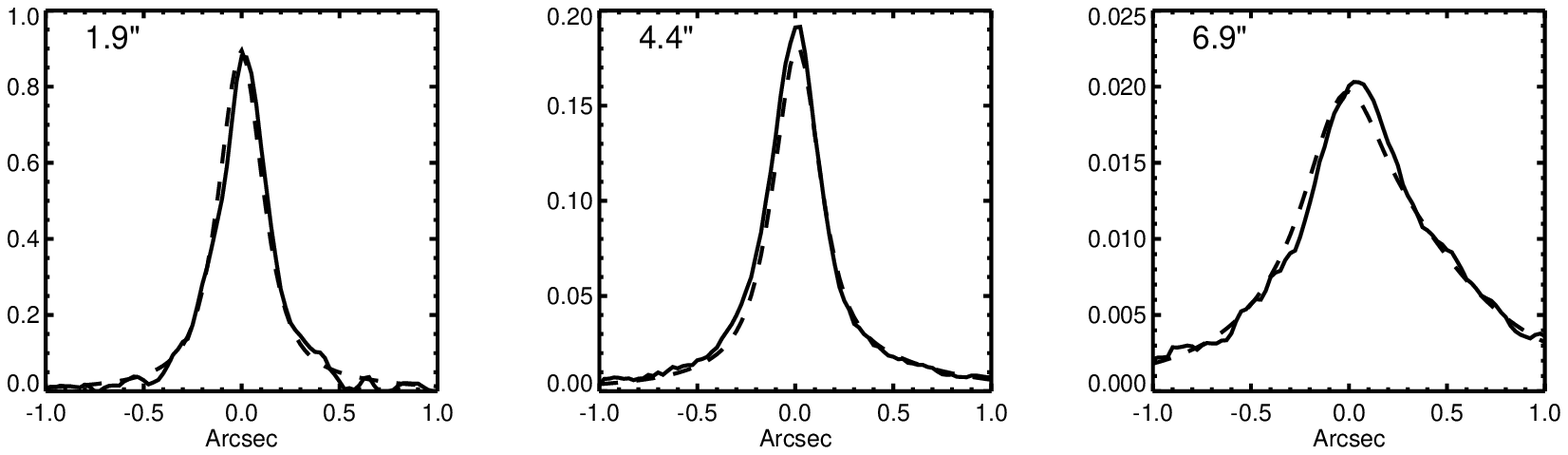}
\caption{AU Mic disk ACS F606W observed and model intensity cross-sections
perpendicular to the midplane at various radii.  The solid lines are the SE
disk, dashed are from the model.  The NE side of the disk is toward the
right.} 
\label{fig11}
\end{figure}

There is likely some tradeoff between the variation in brightness with radius
due to the scattering phase function and the radial density falloff as
characterized by $\alpha$.  A tighter constraint on the phase function would be
possible if the disk could be seen closer to the star than is allowed with the
occulting spot.  The multizonal nature of the disk has been somewhat
artificially decomposed into a set of separate power law descriptions that only
provide constraints on the continuity of the density variation within a zone,
without a rigorous physical basis for them.  Given the number of free
parameters, it is not possible to derive very robust values for both $g$ and
$\alpha$ in the case of a multizonal, edge-on disk.  This would not be a
problem if the disk could be well-resolved in thermal emission, where phase
effects are not present.

\subsection{The Disk in Different Filters}

The disk appears generally similar over the wavelength range of the ACS images.
However, the sharp, thin midplane (FWHM $\approx$ 200 mas) is blurred by the
wavelength-dependent PSF (FWHM $\approx$ 50 -- 72 mas), so the apparent
vertical profile of an intrinsically wavelength-independent disk would appear
to differ among the filters.  Consequently, measurements of color gradients are
complicated.  To examine what sort of PSF-induced variations might be expected,
copies of Model 2 were convolved with Tiny Tim model PSFs for each filter.
These show that for a wavelength-independent disk and illuminating source the
peak midplane intensity ratios at 30 AU are F435W/F606W = 1.05 and F435W/F814W
= 1.16, indicating that the midplane becomes more blurred at longer wavelengths
due to the PSF.  Conversely, more light is redistributed into the wings of the
profile at longer wavelengths.  At $z > $ 0\farcs 4 above the midplane, the
disk model surface brightness is up to 60\% greater in F814W than F435W.  When
the observed images are normalized by the total disk flux in each filter
(measured using the apertures and aperture corrections described below), the
peak midplane brightnesses are equal to those predicted within the measurement
errors, indicating that there are no significant intrinsic color variations
with disk height.

The overall disk color was derived by computing fluxes within regions that
encompass most of the disk's light.  The PSF-convolved models show that by
summing the flux within a 4\farcs 3 long by 0\farcs 9 high box aligned along
the midplane and centered at $r$ = 4\farcs 2, the total flux in F435W would be
1.02$\times$ greater than in F606W and 1.09$\times$ greater than in F814W for a
neutral disk and central source.  These dimensions were chosen based on the
region of significant signal in the observed F435W image.  The same box applied
to each side of the disk in the observed images indicates that the disk is blue
relative to the star.  After aperture corrections derived from the models, the
F435W/F606W flux ratios relative to the star are 1.13 (NW) and 1.15 (SE) and
for F435W/F814W they are 1.35 (NW) and 1.44 (SE).  The estimated measurement
error for this boxsize is 3\%, so it appears that the disk may be slightly more
blue on the SE side.  The SE/NW brightness asymmetry appears to decline toward
longer wavelengths.  The SE/NW brightness ratios measured within the boxes are
1.11 (F435W), 1.09 (F606W), and 1.04 (F814W).  As a consistency check, the flux
ratios in the April and July F606W images agree to within 1\%.

The radial surface brightness profiles for each filter were computed by summing
over 1 $\times$ 35 pixel apertures at each radius centered on the midplane with
aperture corrections derived from the models.  The ratios of these profiles
indicate that the disk becomes increasingly blue at larger radii, at least for
$r =$ 30 -- 60 AU.  On the SE side the F435W/F814W brightness ratio is 1.3 at
30 AU and 1.6 at 60 AU from the star.  On the NW side the change is not as
great, increasing from 1.25 to 1.45 over the same range.  There is no
significant color gradient seen in the F435W/F606W ratios.   Within 30 AU of
the star the disk appears to become more blue toward smaller radii at about
the same rate.  High PSF subtraction residuals in F814W may bias the results
and render this trend suspect.

\subsection{Differences Between Epochs}

AU Mic has a high proper motion of nearly one-half arcsecond per year.  The two
F606W images are separated by over four months, so it is possible to
distinguish background objects from anything that may be co-moving with the
star.  There is nothing in the images except for the disk that appears to be
associated with AU Mic.  Both the galaxy seen along the midplane and the star
SW of the disk moved relative to the disk by an amount and direction consistent
with the proper motion of AU Mic.

\section{Discussion}

\subsection{Indications of an Unseen Perturber?}

The $r < 12$ AU clearing implied by the model fits is consistent with the $r <
17$ AU one predicted by Liu et al. (2004) from the disk's infrared spectral
energy distribution.  Given the low luminosity of the star it cannot be caused
by dust sublimation or expulsion of grains by radiation pressure (KLM04).  As
Liu et al.  suggest, tidal interaction of the dust with an unseen companion
within the clear zone is a strong possibility.  They report no detection of an
infrared companion during a ``shallow'' search using adaptive optics on Keck,
so deeper imaging is certainly warranted.  Given the nearly exact edge-on
orientation of the disk, AU Mic is also a good candidate for detecting
potential planetary transits. 

A low-mass perturber may also explain the asymmetries and localized variations
and the inclination difference between the inner and outer disk.  This latter
appears equivalent to the warping seen in the $\beta$ Pic disk, which has been
explained using hypothetical companions (Augereau et al. 2001).  While the tilt
of the $\beta$ Pic outer disk relative to the inner disk is mostly in the plane
of the sky, in AU Mic's disk it appears inclined mainly along the line of
sight.  Coincidentally, the tilt between $\beta$ Pic's inner and outer disks is
about 3$^{\circ}$, the same as for AU Mic's. 

\subsection{The Vertical Density Distribution}

The AU Mic disk is thin, with an intrinsic FWHM derived from the model fits of
$\sim1.9$ AU for $r < 50$ AU increasing to $\sim3$ AU at $r = 100$ AU.  For
comparison, the $\beta$ Pic disk has a projected FWHM of 15 -- 18 AU between $r
= $ 20 -- 120 AU, and its intrinsic thickness at $r = 100$ AU as derived from
modelling by Kalas \& Jewitt (1995) is $FWHM = $ 7 -- 14 AU.

We stress again that the application of a Lorentzian profile simply provides a
reasonable description of the vertical density distribution, which may be
intrinsically more complex (e.g. the sum of multiple Gaussian distributions of
various scale heights).  It may be possible that the AU Mic disk has a sharper
profile and is more inclined than the fits indicate.  However, a greater
inclination would result in a larger top/bottom asymmetry due to forward
scattering.

\subsection{Disk Color and Grain Properties}

AU Mic is the only known debris disk that appears blue relative to the star.
Other such disks are either largely neutral ($\beta$ Pic; Paresce \& Burrows
1987) or redder than the star (HR 4796, HD 141569A; Schneider et al. 1999,
Clampin et al. 2003) over optical wavelengths.  Those stars, however, are also
A-types, with much greater luminosities.  A potential explanation for the blue
relative color is that the grain-size distribution has a larger proportion of
small ($<1\mu$m) particles than is found in the distributions in the other
disks.  These smaller grains scatter more light at $\lambda = $0.4 $\mu$m than
at 0.9 $\mu$m.  The radiation pressure from A stars could rapidly blow such
particles out of a surrounding disk.  However, an M star like AU Mic exerts 3
-- 5 orders of magnitude less pressure, allowing small grains to remain (Saija
et al. 2003).  Perhaps solar winds may exert more outward pressure on the dust
than radiation in late-type stars. 

The possibility that the outer regions of the disk would be blue relative to
the inner was suggested by KLM04.  In the absence of gas or significant
radiation pressure, Poynting-Robertson drag is the dominant force in altering
the location of collision-produced grains within the disk.  Small grains in the
inner disk should feel the most drag and will migrate quickly into the star,
leaving a relative abundance of larger grains there.   At greater distances
from the star, small grains would encounter less drag and have longer
lifetimes, increasing their proportion relative to the larger grains and thus
increasing the amount of light scattered at shorter wavelengths.

Forward scattering places some constraints on the inclination as greater tilt
would cause the disk on forward side of the midplane to become significantly
brighter than on the other.  The scattering phase parameter value of $g \approx
0.4$ implies that the grain population is different from those found in
optically-thick disks around younger stars such as HH 30 (Burrows et al. 1996)
where $g \approx 0.65$ in the visible and the grain size distribution is
probably similar to the ISM.  The value of $g$ does lie within the 0.3 -- 0.5
range derived for $\beta$ Pic by Kalas \& Jewitt (1995) and is slightly more
than the 0.25 -- 0.35 range of HD 141569A (Clampin et al. 2003).

\section{Conclusions}

The disk of AU Mic has been imaged in scattered light at high resolution with
{\it HST} in {\it BVI}-like passbands to within $\sim0\farcs 5$ (5 AU) of the
star.  The surface brightness profiles of the two sides of the disk are
generally similar, especially when compared to those of the $\beta$ Pic disk,
but localized brightness asymmetries are seen, especially in the SE side. The
disk at $r > 100$ AU is $\sim2\times$ brighter in the NW than the SE.  The disk
is blue relative to the star, perhaps indicating a relative surplus of small
particles compared to grain size distributions in neutral or red disks seen
around earlier-type stars.  The radiation pressure from a late-type star like
AU Mic may be insufficient to force these small grains out of the disk.  An
extended feature at $r = 96$ AU on the NW midplane is a background galaxy.  

Modelling indicates that the inner ($r < 50$ AU) region is seen very close to
edge-on ($\theta < 1^{\circ}$) while the outer is inclined $\sim3^{\circ}$
toward us.  The disk appears to be clear of material within 12 AU of the star,
consistent with the infrared SED.  The disk has a fairly sharp midplane with an
intrinsic vertical FWHM of $<1$ AU at $r < 50$ AU which increases to $\sim3$ AU
at $r = 100$ AU.  The degree of forward scattering by the grains is comparable
to that seen in other debris disks, namely $\beta$ Pic.

\section{Acknowledgements} 

The ACS instrument was developed under NASA contract NAS5-32865, and this
research was supported by NASA grant NAG5-7697.

\end{document}